\documentclass[%
 aip,
 apl,
 amsmath,amssymb,
 numerical,
 reprint,%
]{revtex4-2}
\usepackage{booktabs,multirow,array}
\usepackage[table]{xcolor} 
\usepackage{graphicx}
\usepackage{dcolumn}
\usepackage{bm}
\usepackage{hyperref}

\RequirePackage[normalem]{ulem}
\RequirePackage{color}\definecolor{RED}{rgb}{1,0,0}\definecolor{BLUE}{rgb}{0,0,1} 
\providecommand{\DIFaddbegin}{} 
\providecommand{\DIFaddend}{} 
\providecommand{\DIFdelbegin}{} 
\providecommand{\DIFdelend}{} 
\providecommand{\DIFaddbeginFL}{} 
\providecommand{\DIFaddendFL}{} 
\providecommand{\DIFdelbeginFL}{} 
\providecommand{\DIFdelendFL}{} 
\newcommand{\DIFscaledelfig}{0.5}
\RequirePackage{settobox} 
\RequirePackage{letltxmacro} 
\newsavebox{\DIFdelgraphicsbox} 
\newlength{\DIFdelgraphicswidth} 
\newlength{\DIFdelgraphicsheight} 
\LetLtxMacro{\DIFOincludegraphics}{\includegraphics} 
\newcommand{\DIFaddincludegraphics}[2][]{{\color{blue}\fbox{\DIFOincludegraphics[#1]{#2}}}} 
\newcommand{\DIFdelincludegraphics}[2][]{
\sbox{\DIFdelgraphicsbox}{\DIFOincludegraphics[#1]{#2}}
\settoboxwidth{\DIFdelgraphicswidth}{\DIFdelgraphicsbox} 
\settoboxtotalheight{\DIFdelgraphicsheight}{\DIFdelgraphicsbox} 
\scalebox{\DIFscaledelfig}{
\parbox[b]{\DIFdelgraphicswidth}{\usebox{\DIFdelgraphicsbox}\\[-\baselineskip] \rule{\DIFdelgraphicswidth}{0em}}\llap{\resizebox{\DIFdelgraphicswidth}{\DIFdelgraphicsheight}{
\setlength{\unitlength}{\DIFdelgraphicswidth}
\begin{picture}(1,1)
\thicklines\linethickness{2pt} 
{\color[rgb]{1,0,0}\put(0,0){\framebox(1,1){}}}
{\color[rgb]{1,0,0}\put(0,0){\line( 1,1){1}}}
{\color[rgb]{1,0,0}\put(0,1){\line(1,-1){1}}}
\end{picture}
}\hspace*{3pt}}} 
} 
\LetLtxMacro{\DIFOaddbegin}{\DIFaddbegin} 
\LetLtxMacro{\DIFOaddend}{\DIFaddend} 
\LetLtxMacro{\DIFOdelbegin}{\DIFdelbegin} 
\LetLtxMacro{\DIFOdelend}{\DIFdelend} 
\DeclareRobustCommand{\DIFaddbegin}{\DIFOaddbegin \let\includegraphics\DIFaddincludegraphics} 
\DeclareRobustCommand{\DIFaddend}{\DIFOaddend \let\includegraphics\DIFOincludegraphics} 
\DeclareRobustCommand{\DIFdelbegin}{\DIFOdelbegin \let\includegraphics\DIFdelincludegraphics} 
\DeclareRobustCommand{\DIFdelend}{\DIFOaddend \let\includegraphics\DIFOincludegraphics} 
\LetLtxMacro{\DIFOaddbeginFL}{\DIFaddbeginFL} 
\LetLtxMacro{\DIFOaddendFL}{\DIFaddendFL} 
\LetLtxMacro{\DIFOdelbeginFL}{\DIFdelbeginFL} 
\LetLtxMacro{\DIFOdelendFL}{\DIFdelendFL} 
\DeclareRobustCommand{\DIFaddbeginFL}{\DIFOaddbeginFL \let\includegraphics\DIFaddincludegraphics} 
\DeclareRobustCommand{\DIFaddendFL}{\DIFOaddendFL \let\includegraphics\DIFOincludegraphics} 
\DeclareRobustCommand{\DIFdelbeginFL}{\DIFOdelbeginFL \let\includegraphics\DIFdelincludegraphics} 
\DeclareRobustCommand{\DIFdelendFL}{\DIFOaddendFL \let\includegraphics\DIFOincludegraphics} 
\usepackage[utf8]{inputenc}
\usepackage[T1]{fontenc}
\usepackage{mathptmx}
\usepackage{siunitx}
\usepackage{booktabs}
\usepackage{xcolor}
\usepackage{siunitx}
\usepackage{appendix}

\graphicspath{{gfx}}

\begin{document}

\preprint{AIP/123-QED}

\title[3D surface profilometry using neutral helium atoms]{3D surface profilometry using neutral helium atoms}

\author{Aleksandar Radi\'{c}}
\email[Author to whom correspondence should be addressed: ]{ar2071@cam.ac.uk}
\affiliation{Department of Physics, Cavendish Laboratory, 19 J.J. Thomson Avenue, University of Cambridge, Cambridge, CB3 0HE, UK}

\author{Sam M. Lambrick}
\affiliation{Department of Physics, Cavendish Laboratory, 19 J.J. Thomson Avenue, University of Cambridge, Cambridge, CB3 0HE, UK}
\affiliation{ 
Ionoptika Ltd, Units B5-B6, Millbrook Close, Chandlers Ford, Southampton, S053 4BZ, UK}

\author{Nick A. von Jeinsen}
\affiliation{Department of Physics, Cavendish Laboratory, 19 J.J. Thomson Avenue, University of Cambridge, Cambridge, CB3 0HE, UK}

\author{Andrew P. Jardine}
\affiliation{Department of Physics, Cavendish Laboratory, 19 J.J. Thomson Avenue, University of Cambridge, Cambridge, CB3 0HE, UK}

\author{David J. Ward}
\affiliation{Department of Physics, Cavendish Laboratory, 19 J.J. Thomson Avenue, University of Cambridge, Cambridge, CB3 0HE, UK}
\affiliation{ 
Ionoptika Ltd, Units B5-B6, Millbrook Close, Chandlers Ford, Southampton, S053 4BZ, UK}

\date{\today}

\begin{abstract}
Three-dimensional mapping of surface structures is important in a wide range of biological, technological, healthcare and research applications. Neutral helium atom beams have been established as a sensitive probe of topography and have already enabled structural information to be obtained from delicate samples where conventional probes would cause damage. Here, we empirically demonstrate, for the first time, a reconstruction of a complete surface profile using measurements from a modified scanning helium microscope (SHeM), using the heliometric stereo method and a single detector instrument geometry. Results for the surface profile of tetrahedral aluminum potassium sulphate crystals demonstrate the areas of surfaces and facet orientations can be recovered to within $5\%$ of the expected values.
\end{abstract}

\maketitle

\noindent Accurate measurements of surface topography (often termed surface metrology) is crucial to modern research and development, with widespread applications from understanding material behavior to optimizing manufacturing\cite{jiang_paradigm_2007,jiang_paradigm_2007-1}.
Significantly, surface metrology can also provide assurance that products and materials meet required standards for functionality, quality, and safety\cite{Mathia2011}. As new technologies emerge, for example in micro- and nano-scale organic and quantum devices, the reliable measurement of surface topography on an ever smaller scale is crucial. Scanning helium microscopy (SHeM)\cite{Witham2011,Barr2014,Koch2008,palau_neutral_2023,flatabø2023reflection} is an emerging technique that uses a narrow beam of neutral helium atoms ($\geq\SI{300}{nm}$\cite{witham_increased_2012}), generated \textit{via} a collimating pinhole to produce topographic micrographs; in the current work the instrument is of the type presented by Barr et al.\cite{Barr2014}. The use of a neutral beam has two key advantages for performing topographic measurements, which taken together are unique. First, it enables the use of a very low energy beam without the resolution being diffraction limited: the typical beams used in SHeM offer a de Broglie wavelength of  $\SI{0.06}{nm}$, corresponding to energies of  $\SI{64}{meV}$\cite{Holst2021,lifPaper}, sufficiently small to render beam damage, as observed with charged beam methods\cite{Ramachandra2009}, impossible. Second, thermal helium atoms scatter off the outermost electron density distribution, located $2-\SI{3}{\angstrom}$ above the ion cores of surface atoms, which results in information exclusively from the surface, without any contribution due to beam penetration into the bulk\cite{Holst2021}. Existing scattering techniques for performing surface metrology, such as optical profilometry, laser scanning confocal microscopy, scanning electron microscopy, interferometric profilometry and structured light scanning\cite{Kournetas2017,Arvidsson2006,Valverde2013}, either use highly energetic probe particles, which can induce changes in the sample, or can have a non-negligible interaction volume with the surface and thus do not measure the surface topography exclusively. Contact techniques, such as stylus profilometers, scanning tunneling microscopes and atomic force microscopes\cite{Al-Nawas2001,Rupp2003,Svanborg2010,Wennerberg2014}, provide a more direct surface profile measurement through contact with a physical probe. However these are largely limited in either lateral resolution (stylus profilometers) or in aspect ratio (AFM/STM), can induce changes in the sample due to the direct contact with the probe, and in all cases convolution with the probe tip must be considered.  

SHeM has demonstrated its ability to measure surface topography qualitatively\cite{Fahy2018,Lambrick2018,fahy_highly_2015,pranav}, and two previous publications have shown SHeM to be capable of providing individual pieces of 3D information. Myles et al.\cite{Myles2019} applied stereophotogrammetry to taxonomy, measuring specific dimensions of biological specimens.  Lambrick et al.\cite{LambrickMultiple2020} demonstrated quantitative 3D measurements can be made by interpreting specific contrast features of SHeM to precisely depth profile trenches in silicon. Both approaches make accurate measurements of specific dimensions, but do not yield a 3D view of the entire sample. In the current work we empirically demonstrate an alternative approach, that has recently been proposed theoretically\cite{Lambrick3d}, \emph{heliometric stereo}, that can obtain complete topographic maps using multiple helium micrographs acquired with different detection conditions. We apply the method using the Cambridge SHeM\cite{Barr2014}, a single detector instrument, showing the approach can be utilised with the current generation of SHeM instruments, with a normally incident beam instead of the previously published $\ang{45}$ angle of incidence. The modification of our SHeM to operate in normal incidence was made possible by the use of a modular, 3D printed pinhole plate optical element\cite{m_bergin_complex_2021}. Although SHeM micrographs using normal incidence have been published previously \cite{LambrickDiffuse2022}, the current work presents the novel pinhole plate geometry, and its effects on imaging, in detail. Further, the success of the reconstruction validates recent reports that the scattering of atoms from disordered surfaces is almost purely diffuse\cite{LambrickDiffuse2022}, an important result for the development in the field of SHeM.

Heliometric stereo applies an optical technique, photometric stereo\cite{woodham_photometric_1980}, to SHeM. A full discussion of the equivalences and differences between image formation in SHeM and optical imaging is given alongside the theoretical foundations of heliometric stereo by Lambrick \& Palau et al.\cite{Lambrick3d}. Here, we summarize the key points that are relevant to the current work. The method relies upon the assumption that a given point on a surface will scatter the incident helium beam with a known scattering distribution, formally termed the bidirectional reflection distribution function (BRDF)\cite{StoverOptics}. The distribution defines the scattered intensity as a function of the incoming and outgoing angles of the helium beam relative to the surface. If the microscope detector is fixed at a known angle to the surface and the incident direction is known, then the recorded intensity is solely a function of surface orientation and scattering distribution. SHeM instruments have a well defined scattering geometry\cite{LambrickDiffuse2022}, which gives a direct relationship between the intensity of each pixel in a micrograph and the local surface orientation of the corresponding point on the sample.

Recent experimental results have shown that the vast majority of surfaces studied in SHeM exhibit almost exclusively diffuse scattering \cite{LambrickDiffuse2022,LambrickMultiple2020,Lambrick2018,Fahy2018,hatchwell2023measuring}.  These materials are typically described as `technological samples' or `unprepared surfaces', in contrast to surface-science studies where surfaces are prepared and maintained in a pristine, atomically perfect state, and where specular or Bragg scattering is often dominant\cite{Holst2021,lifPaper,estermann_beugung_1930,holst_atom-focusing_1997,corem_ordered_2013}. The observed diffuse distribution is consistent with Knudsen's cosine law\cite{Knudsen1935}, and is analogous to Lambert's cosine law for visible light\cite{Lambert1760}. The observation that diffuse scattering is the dominant scattering process for `technological' surfaces can be understood as a result of the de Broglie wavelength of the helium atoms: $\lambda\sim\SI{1}{\angstrom}$ being comparable to the inter atomic spacing in solids. Thus the scattered atoms are sensitive to surface disorder on any length scale between the width of the beam ($\sim\SI{5}{\micro\metre}$ in the current work, but down to $\sim\SI{300}{\nano\metre}$ has been reported\cite{witham_increased_2012}), down to the atomic scale which is commensurate with the wavelength of the beam. Most `technological' samples will have disorder on one or more of the length-scales between the beam spot size and the atomic scale and thus will exhibit randomized, or diffuse scattering.  

\begin{figure}[b]
    \centering

    \textsf{\small\input{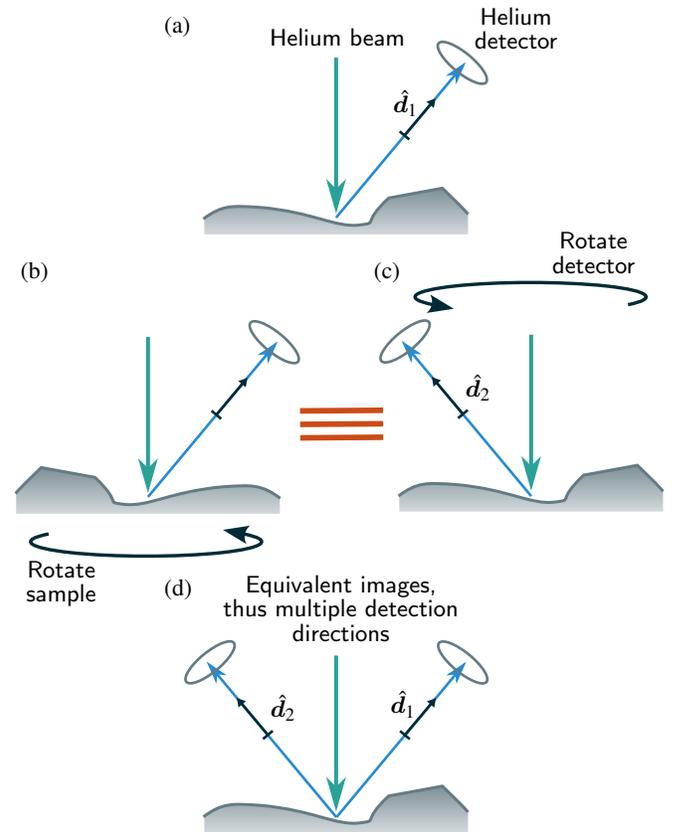}}
    
    \caption{Illustration of multiple image acquisition with a single detector instrument; (a) a helium micrograph is obtained. (b) the sample is rotated about the helium beam axis to obtain a second micrograph. (c) the rotation of the sample is equivalent to a rotation of the detector. (d) We now have 2 micrographs with 2 different detection directions, corresponding to 2 effective detectors. The images with multiple detection directions can now be used to perform a heliometric stereo reconstruction. Note that a rotation of the sample about an axis other than the helium beam axis would result in a different illumination of the sample and therefore the correspondence with a rotation of the detector would be lost.}
    \label{fig:rotating_detectors}
\end{figure}

The following derivation is a summary of the heliometric stereo method previously presented by Lambrick \& Palau et al.\cite{Lambrick3d} Mathematically, diffuse scattering follows a cosine distribution centered on the surface normal,

\begin{equation}\label{eq:cosine}
    I(\theta) = \rho\cos\theta = \rho\hat{\bm{n}}\cdot\hat{\bm{d}},
\end{equation}

where $\theta$ is the angle between the local surface normal, $\hat{\bm{n}}$, and the outgoing scattering direction $\hat{\bm{d}}$. It can be shown that for cosine scattering with a macroscopic circular detection area -- such as is present experimentally in SHeM\cite{Barr2014} -- that equation \ref{eq:cosine} holds the same mathematical form up to a multiplicative constant\cite{Lambrick3d}. An equation with identical form exists for light in photometric stereo, however, due to differences in image projection for light, $\hat{\bm{d}}$ represents the illumination direction; both methods then follow from equation \ref{eq:cosine} with different definitions of $\hat{\bm{d}}$. As the scattering distribution directly relates the intensity scattered in a particular direction to the orientation of the surface, it is possible to acquire the surface orientation from a series of measured intensities. The orientation of a surface is defined by two angles, plus we allow for a constant of proportionality, the albedo factor ($\rho$ in equation \ref{eq:cosine}), hence there are 3 degrees of freedom. Therefore a minimum of 3 data points are needed to solve the problem. In practice SHeM data is often signal-to-noise ratio limited and there may be minor deviations to the cosine law, thus at least 4 are recommended for a reconstruction to be robust. By having multiple detection directions in a SHeM system, equation \ref{eq:cosine} may be written in matrix form,
\begin{equation}\label{eq:ch3D:basic_photostereo}
	\Vec{I}_{(x',y')}=\rho \mathsf{D}\hat{\bm{n}},
\end{equation}
where $\Vec{I}$ is a \(m\)-dimensional vector of pixel intensities, all for the same position on the sample, corresponding to \(m\) images taken with different detection directions. $\mathsf{D}$ is a \(m \times 3\) matrix containing the normalized vectors pointing from the scattering point to the detectors. The linear system is thus solved,
\begin{gather}
	\rho_{(x^\prime,y^\prime)} = |\mathsf{D}^{-1}\Vec{I}_{(x^\prime,y^\prime)}|,\label{eq:ch3D:solvePhotostereo1}\\
	\hat{\bm{n}}_{(x^\prime,y^\prime)} = \frac{1}{\rho_{(x^\prime,y^\prime)}}\mathsf{D}^{-1}\Vec{I}_{(x^\prime,y^\prime)}.\label{eq:ch3D:solvePhotostereo2}
\end{gather}
Assuming the height of the surface can be described by a continuous function of the lateral position, {\em i.e.} $z = f(x, y)$, then
\begin{gather}
	\hat{\bm{n}}(x,y) = \bm{\nabla} F(x,y,z) = \bm{\nabla}[z- f(x,y)].\label{eq:ch3D:gradient_field}
\end{gather}
Thus once the surface normals are found, the gradient field given by equation \ref{eq:ch3D:gradient_field} may be integrated to obtain an equation $z=f(x,y)$ of the surface profile, \emph{i.e.} a topographic map of the sample. A regularized least squares approach, developed by Harker and O'leary\cite{HarkerOLeary2008,HarkerMatlab}, is used in the current work. The method has proved robust to noise in simulated SHeM micrographs\cite{Lambrick3d}.

\begin{figure}
	\centering
	\textsf{\small \input{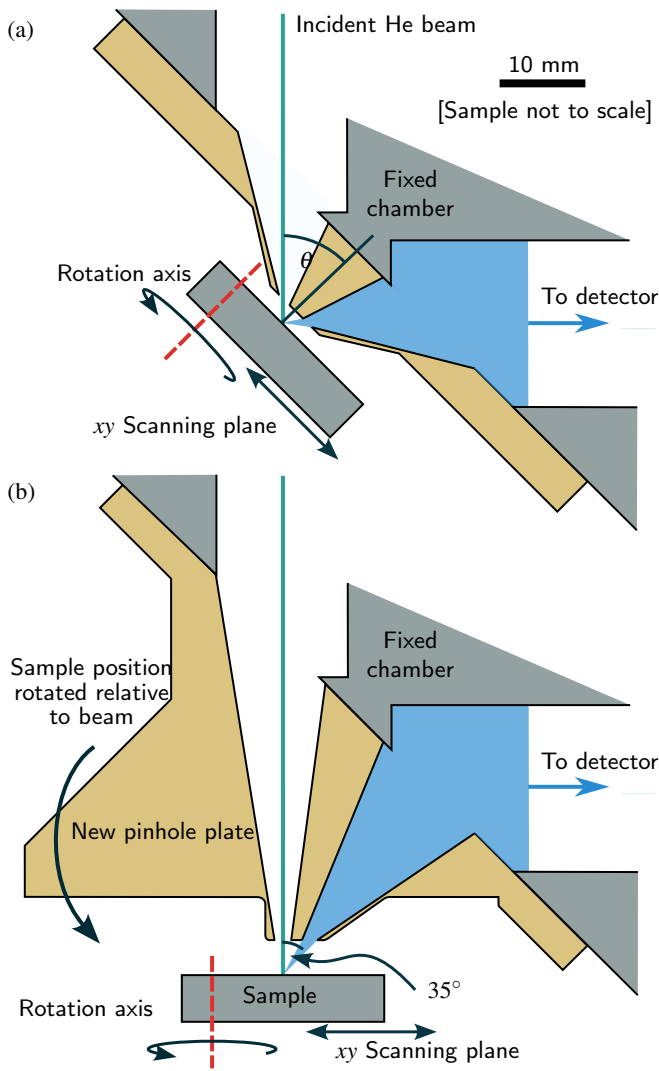}}

    \caption{Cross sectional diagram of the sample manipulation and pinhole plate of the Cambridge SHeM. (a) with the rotation stage introduced to the original, $\theta=\ang{45}$, configuration, as used by von Jeinsen et al.\cite{lifPaper}. (b) with the new pinhole plate macroscopically rotating the system to operate in a normal incidence configuration. Highlighted in yellow are the modular pinhole plates, as distinct from the fixed vaccuum chamber walls. The new design adds a small, but largely negligible, distance of $\sim\SI{1.5}{\centi\metre}$ to the overall pathway to the ionization volume of the detector, which is $\sim\SI{60}{\centi\metre}$.}
	\label{fig:pinhole_plate}
\end{figure}

The two key assumptions heliometric stereo relies on are: (i) the direct relationship between surface orientation and detected signal expressed in equation \ref{eq:cosine}, and (ii) the assumption of a surface that can be described as a continuous function, $z=f(x,y)$. Requirement (ii) makes the method suitable for samples with continuous changes in surface topography, \emph{i.e.} without overhangs or large vertical faces. Requirement (i) is broken by some identified contrast features in SHeM, multiple scattering\cite{LambrickMultiple2020} and masking\cite{Fahy2018,Lambrick2018}, which are most prominent near vertical edges, but can also be reduced by careful consideration of the scattering geometry. Simulated results from our previous work demonstrate that modest regions of multiple scattering or masking will only cause local errors in the reconstruction and will not prevent a good reconstruction overall.

A key result found in previous work\cite{Lambrick3d} was that \emph{rotations about the beam axis} in SHeM may be used to create multiple `effective detection directions', on an instrument with a single detector. In figure \ref{fig:rotating_detectors} that acquisition process is illustrated: an image is taken with the sample in one orientation (a), then the sample is rotated and a second image is taken (b). As the rotation occurred around the beam axis, the rotation of the sample is equivalent to a rotation of the detector (c). Therefore multiple detection directions may be sampled in an instrument with a single physical detector (d). As discussed in the theoretical development of heliometric stereo\cite{Lambrick3d}, if the rotation is performed about an axis other than one parallel to the beam axis the illumination of the sample is changed, and thus the correspondence between sample rotations and multiple detectors is lost.

\begin{figure}[t]
    \centering
    \includegraphics[width=0.55\linewidth]{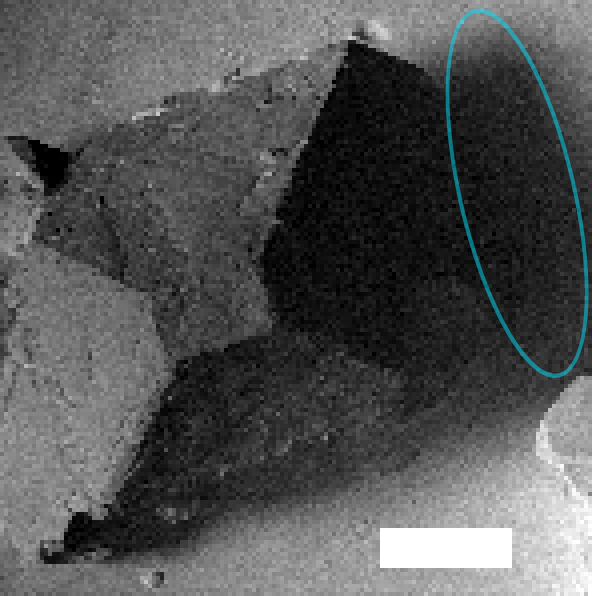}
    \caption{SHeM micrograph of the octahedral reconstruction candidate crystal with $\SI{500}{\micro \metre}$ scale bar. 
    The dark region highlighted is an example of a helium mask, where the line of sight between the sample and detector is blocked.}
    \label{fig:crystals_overview}
\end{figure}

\begin{figure*}
    \centering
    \includegraphics[width=0.95\linewidth]{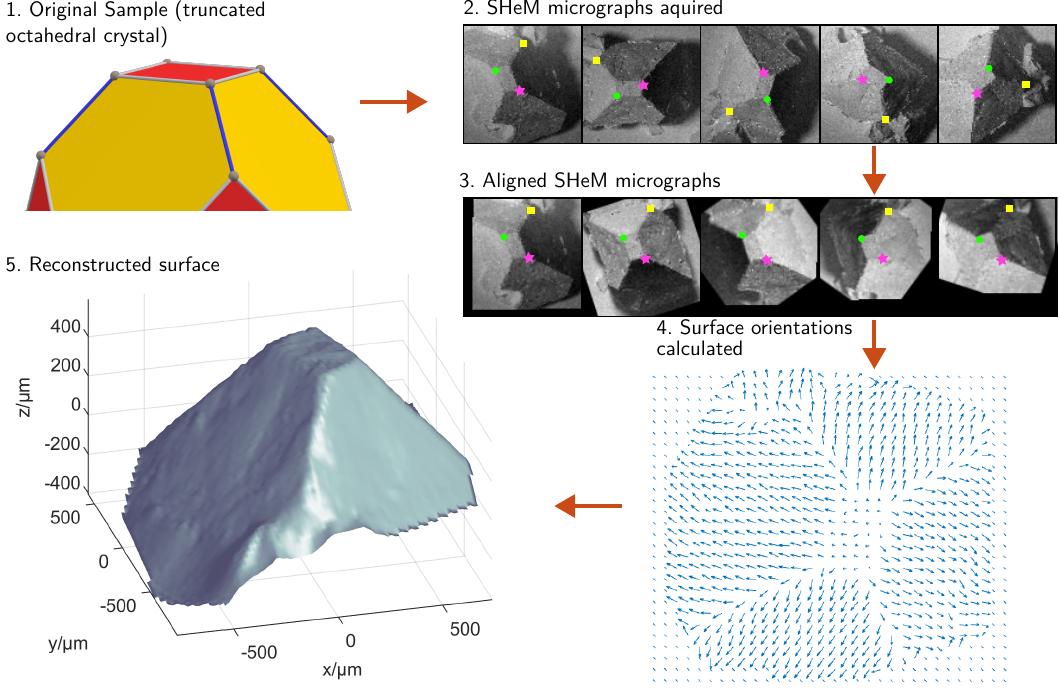}
    \caption{Experimental procedure of performing heliometric stereo. (1)-(2) A series of SHeM micrographs are taken of the sample. (2)-(3) Reference points are used to align the micrographs. (4) the detected helium intensities from the aligned micrographs are used with equation \ref{eq:ch3D:solvePhotostereo2} to give surface normal vectors. (5) the normal vectors are integrated to a give a surface topography map.\newline\footnotesize{Panel 1 adapted from a figure by T. Piesk, CC BY 4.0, published 2018, https://commons.wikimedia.org/w/index.php?curid=66347083.}}
    \label{fig:BigFigure}
\end{figure*}

The Cambridge SHeM was designed to operate with an incidence angle of $\ang{45}$\cite{Barr2014} allowing for topographic imaging as well as studies of any specular signal. In the current work we modified the Cambridge SHeM to operate in a normal incidence configuration through a new 3D printed `pinhole plate' optical element\cite{m_bergin_complex_2021,radicPlastics}. The pinhole plate mounts the collimating pinhole, defines the detection geometry and, in the Cambridge SHeM, is used to mount the sample and sample manipulation stages. The old (a) and new (b) configurations are shown in figure \ref{fig:pinhole_plate}, with the incident helium beam and rotation axis highlighted. The new pinhole plate macroscopically changes the mounted orientation of both the sample and the sample manipulation stages relative to the incident beam: in figure \ref{fig:pinhole_plate} (b) the beam is now incident at $\ang{0}$ to the sample and the axis of rotation is parallel to the beam. However, while the rotation axis and the beam are parallel, they are not coaxial, thus an alignment procedure was needed to ensure that the reconstruction candidate remained centered in the micrographs as the sample was rotated. The same rotation and alignment procedure as used by von Jeinsen et al.\cite{lifPaper} was applied here. As seen in figure \ref{fig:pinhole_plate} the new design does introduce a change in the path from the collection aperture to the detector itself, the change in the length of the path is, however, small ($\sim3\%$). While the exact shape of the path after the initial cone has been shown in previous work not to affect contrast observed in SHeM micrographs\cite{Lambrick2018}.

The contrast feature of masking occurs when the line of sight between the sample and detector is blocked and results in dark `shadow-like' features appearing in micrographs -- an example is highlighted in figure \ref{fig:crystals_overview}. These dark regions break the mathematical relationship between pixel intensity and surface orientation expressed in equation \ref{eq:cosine}, thus it is important to minimize the region of sample under mask for heliometric stereo. The angle between the incident helium beam and the detection direction was therefore kept reasonably small, $\sim\ang{35}$ compared to $\ang{90}$ in the original configuration. In addition, the detector aperture was shaped to occupy a circular region of solid angle, ensuring the $I\propto\cos\theta$ relation holds -- see appendix A of Lambrick \& Palau et al.\cite{Lambrick3d} for details. We note that our approach of changing only the key optical element, i.e. the pinhole plate, to alter the imaging geometry of the SHeM could be used to achieve other goals.  For example, changing the incident and detection angles could highlight either certain contrast features or could change the image projection, potentially also allowing an improvement in the resolution of the instrument.

To test the reconstruction accuracy, a sample of well defined topography was needed that matched the basic assumptions of heliometric stereo, principally a continuous $z=f(x,y)$ surface. Aluminum potassium sulphate crystals ($KAl(SO_{4})_{2}. 12 H_{2}O$) were chosen. They posses a well defined octahedral structure and can be readily grown to form single crystals on micron to millimeter length-scales. Figure \ref{fig:crystals_overview} shows an overview SHeM micrograph of the aluminum potassium sulphate crystal selected for imaging.

To demonstrate topographic reconstruction we acquired helium micrographs at 5 equally spaced azimuthal intervals of $\ang{72}$. A slight overconstraint of the system in equation \ref{eq:ch3D:solvePhotostereo2} was used to account for noisy data or for slight deviations from ideal diffuse scattering. In addition, the rotational stage of the sample manipulator has a small amount of mechanical inaccuracy which limits the accuracy to which the center of rotation can be determined to a few pixels. Consequently the tracking of the sample through rotations was imperfect, taking more micrographs minimised the image area where fewer than three measurements were present. We note that improvement of rotational tracking will be essential in future heliometric stereo reconstruction in single detector SHeM. 

The reconstruction process is illustrated in figure \ref{fig:BigFigure} and starts by manually identifying common points across the micrographs to correlate and align the separate azimuthal images. Distinct points on the sample were selected that are accurately identifiable across micrographs, as shown in panel 2.  A transformation matrix, using only translational and rotational components, was calculated to create a mapping of each micrograph onto the same axis. Mapping to a common set of axes is crucial because the pixels in each image must correspond to the same point on the sample surface so that surface orientation may be inferred. Panel 3 shows the images after alignment with 3 tracking points highlighted. Where at least 3 micrographs overlapped, the intensities from the micrographs were used to calculate the surface normals according to equation \ref{eq:ch3D:solvePhotostereo2}.  Normal vectors for each pixel were generated, as shown in Figure \ref{fig:BigFigure} panel 4. Pixels which cannot be correlated across at least 3 micrographs are assigned a surface orientation parallel to the $z$ axis because otherwise the linear system, shown in equation \ref{eq:ch3D:basic_photostereo}, becomes under-determined. Orientation parallel to the $z$ axis was chosen because it is the average of all possible orientations It is necessary to apply such a condition so that under-determined pixels have minimal effect on the integration of surface normal vectors which gives the final surface reconstruction.  Finally the normal map is integrated using the regularized least squares method\cite{HarkerOLeary2008,HarkerMatlab}, giving the surface reconstruction, shown in panel 5. The reconstruction captures all qualitative aspects of the crystal: we are clearly observing the top half of a truncated octahedron.

Two metrics were employed to assess the quantitative accuracy of the reconstruction. First, the surface area of the top of the crystal was used to ensure the reconstruction process does not distort the size and shape of the sample. The area of the top of the crystal was confirmed both via the 2D SHeM micrographs and independently via SEM. It should be noted that measurements taken directly from 2D SHeM micrographs and SEM images will both be affected by any tilt in the sample, or sample mounting. Second we measured the angles between the facets of the reconstructed surface and compared them to the known facet angles of an octahedral aluminum potassium sulphate crystal.

The rectangular top face of the crystal was measured to have side lengths $0.438\pm\SI{0.009}{mm}$ by $0.250\pm\SI{0.009}{mm}$ giving a surface area $(0.110\pm0.005)\,\mathrm{mm}^2$ in the SEM micrographs, compared with $0.423\pm\SI{0.015}{\milli\metre}$ by $0.248\pm\SI{0.015}{\milli\metre}$ giving a surface area of $(0.105\pm0.007)\,\mathrm{mm}^2$ in the 2D SHeM micrographs. Measurement of the side lengths in the final 3D reconstruction gives $0.424\pm\SI{0.013}{mm}$ by $0.254\pm\SI{0.013}{mm}$ for a surface are $(0.108\pm0.006)\,\mathrm{mm}^2$; all three area values agree within experimental uncertainty. The agreement indicates that there is no significant distortion occurring due to the reconstruction algorithm. Full table of side lengths and calculated area values in pixel and millimeter units are recorded in the Supplementary Information in Section II, table S1.

The angle between the top facet of the crystal and the side facets for a truncated octahedron are $\ang{125}$ (to 3 s.f.). To measure the angle between the reconstructed facets, the surface height going left to right across the crystal and top to bottom were averaged, producing two data sets each crossing the top of a trapezoid. Linear regression was applied to each side of the two trapezoidal plots, resulting in four measurements of the facet angle; $\ang{136}$, $\ang{122}$, $\ang{129}$ and $\ang{131}$, with a mean value of $\ang{129}$ a $5\%$ overestimate compared to the ideal octahedron. The uncertainties in the linear regression were small, therefore the deviations from the expected value are not attributed to random noise. Given an overall high quality reconstruction with small deviation there is scope for a further study into the absolute precision of heliometric stereo for different types of samples.

Our results show that accurate surface profile reconstructions can be acquired using heliometric stereo for samples suited to the technique -- those that conform to the key assumptions of the method, most notably the assumption of a continuous function $z = f(x,y)$. Therefore the base method will struggle with overhangs and vertical faces in samples, features that prove challenging in many other techniques too, although as shown with simulated data small vertical faces only cause local errors in the reconstruction\cite{Lambrick3d}. An advantage of the approach, as now shown both with experimental (current work) and simulated data\cite{Lambrick3d} is that sample heights may be of the same magnitude as the lateral extent of the sample, enabling reconstruction of samples with aspect ratios of unity. With previously reported lateral SHeM spot sizes down to $\SI{300}{nm}$\cite{witham_increased_2012}, existing SHeM instruments could, with minor modification, access sub-optical resolution surface reconstructions for moderate to high aspect ratio samples.


We present the first topographic reconstruction performed using scanning helium microscopy. In particular, we describe the process of performing heliometric stereo on a single detector instrument. The topographic map of the chosen test sample successfully reproduced all qualitative features and demonstrated reconstruction of the shape and size of the structure to within $5\%$ accuracy. The modifications to our single detector SHeM to allow the application of heliometric stereo means that the method can now be readily used, and in principle the changes are applicable to any existing SHeM design. The success of the reconstruction process further validates recent work demonstrating that diffuse scattering is dominant for technological surfaces in SHeM. The principle of implementing a new pinhole plate optical element for normal incidence also holds promise for allowing different modes of operation on a single SHeM instrument. The ease of modification means that design choices for different purposes, for example optimization for higher spatial resolution, could be implemented without rebuilding the whole machine. In the future we anticipate that the methods presented by Myles et al.\cite{Myles2019} or the use of contrast features as presented by Lambrick et al.\cite{LambrickMultiple2020} will be combined with heliometric stereo for higher accuracy surface metrology, especially as SHeM resolution pushes into the nanoscale. The success in reconstruction also highlights the desire for a multiple detector SHeM where heliometric stereo could be applied with shorter measurement times and without the need for image alignment.

\noindent\textit{Supplementary Material.} See Supplementary Material for details and diagram of facet angle measurements of the reconstructed pyramid shown in figure 4. Supplementary Material also contains all surface face measurements (with error values) from 2D SEM, 2D SHeM and 3D reconstruction in table S1, and image parameters used to acquire SHeM images used in reconstruction (pixel size, pixel image dimensions, pixel dwell time and time taken per micrograph acquisition).

\noindent\textit{Data Availability.} A supporting data pack containing both raw and processed data, and Supplementary Material can be found at \url{http://doi.org/10.17863/CAM.107537}.

\noindent\textit{Acknowledgements.} The work was supported by EPSRC grant EP/R008272/1. The authors acknowledge support by the Cambridge Atom Scattering Centre (\url{https://atomscattering.phy.cam.ac.uk}) and EPSRC award EP/T00634X/1. SML acknowledges support from IAA award EP/X525686/1 and funding from Mathworks Ltd. We would like thank Boyao Liu for useful discussions. The authors acknowledge support from Ionoptika Ltd.

\nocite{*}
\bibliographystyle{apsrev4-2}
\bibliography{heliometric_references}

\clearpage










\section{Facet angle measurement}
\begin{figure}[b]
    \centering
    \includegraphics[width=\linewidth]{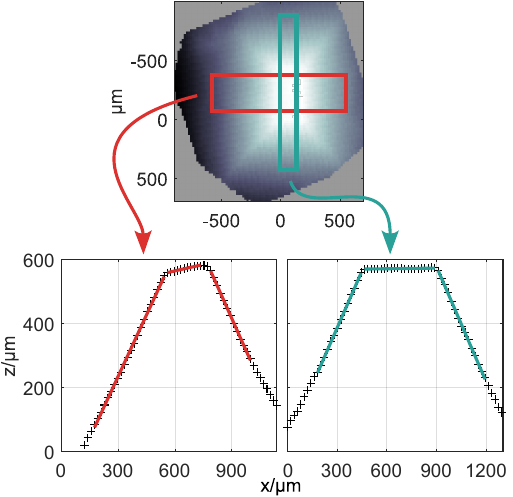}
    \caption{Height map of the reconstructed surface from figure 4 panel 4 with the horizontal and vertical slices used to measure the facet angles highlighted. The data for the slices is averaged with a linear regression performed for each facet, with the resulting gradients, plotted in colour, used to calculate the angles.}
    \label{fig:measurements}
\end{figure}

Figure \ref{fig:measurements} shows a height map of the reconstructed crystal, along with the extracted horizontal and vertical profiles. The linear fits that are used to calculate the measured facet angles are plotted over the extracted data points.

\newpage
\section{Surface face measurement}\label{sec:surface_face_measurement}

\begin{table}[h!]
\centering
    \begin{tabular}{@{} l*{4}{>{$}c<{$}} @{}}
    \toprule
     & \multicolumn{4}{c@{}}{Dimensions of Pyramid Truncated Face}\\
    \cmidrule(l){2-5}
     & Long & Short & Area & Units\\
    \midrule
    \multirow{3}{*}{SEM}  &240.5\pm 5 & 137.2\pm5 & \num{3.30e4}\pm 25 & px\\
    & 0.438\pm 0.009 & 0.250\pm 0.009 & 0.110 \pm 0.005 & mm  \\
    
    \multirow{3}{*}{}\\
    
    \multirow{3}{*}{2D SHeM}  & 28.2\pm 1 & 16.5\pm 1 & 466 \pm 1 & px\\ 
     & 0.423\pm 0.015 & 0.248 \pm 0.015 & 0.105 \pm 0.007 & mm\\
    
    \multirow{3}{*}{}\\
    
    \multirow{3}{*}{3D SHeM}  & 154.9\pm 1 & 92.6 \pm 1 & \num{1.43e3} \pm 1 & px\\ 
    & 0.424 \pm 0.013 & 0.254 \pm 0.013 & 0.108 \pm 0.006 & mm\\

    \bottomrule
    \end{tabular}
    \caption{Measured side lengths, and calculated areas, of the top face of the truncated pyramid in figures 3 and 4 used to evaluate heliometric stereo reconstruction accuracy. Pixel to millimeter conversions, SEM: $\SI{1}{\milli\metre} = 549\pm\SI{2}{px}$, 2D SHeM: $\SI{1}{\milli\metre} = 66.7\pm\SI{1}{px}$, 3D SHeM: $\SI{1}{\milli\metre} = 365\pm\SI{1}{px}$}
    \label{tab:pyramid_measurements}
\end{table}

\vspace{0.5em}
\section{Image parameters}

The acquisition time for micrographs in figure 4 was 3h 24min per micrograph. Pixel sizes of $\SI{20}{\micro\metre}$ were used with a dwell time of $\SI{750}{\milli s}$. $101\times101$ pixels were used for the micrograph.

The acquisition time for the micrograph in figure 3 was 7h 16min. Pixel sizes of $\SI{15}{\micro\metre}$ were used with a dwell time of $\SI{495}{\milli s}$. $167\times167$ pixels were acquired -- the version presented is cropped.


\end{document}